\documentclass{article}
\usepackage[noend]{algorithmic}
\usepackage[ruled]{algorithm2e}
\usepackage{amsmath} 
\usepackage{pgf}
\usepackage{tikz}
\usetikzlibrary{arrows,automata}
\usepackage[latin1]{inputenc}
\usepackage{verbatim}
\usepackage{graphicx} 
\usepackage{subfigure,tikz}
\usepackage{authblk}
\begin{document}
\title{{\bf A Simple Algorithm for Global Value Numbering}}
\author{{\normalsize Nabizath Saleena and Vineeth Paleri}}
\affil{{\normalsize Department of Computer Science and Engineering\\National Institute of Technology Calicut, India.}\\  \texttt{{\small {\{saleena,vpaleri\}@nitc.ac.in}}}}
\date{}
\maketitle
\begin{abstract}
Global Value Numbering(GVN) is a method for detecting redundant computations in programs. Here, we introduce the problem of Global Value Numbering in its original form, as conceived by Kildall(1973), and present an algorithm which is a simpler variant of Kildall's. The algorithm uses the concept of \textit{value expression} - an abstraction of a set of expressions - enabling a representation of the equivalence information which is compact and simple to manipulate.  
\end{abstract}

\maketitle

\section{Introduction}
Detection and elimination of redundant computations have been interesting topics in the area of code optimization in compilers. Value Numbering originated as a method for detecting redundant computations within a basic block (known as Local Value Numbering). The basic idea is to assign a number to each expression in such a way that \textit{equivalent} expressions are assigned the same number  \cite{appel}. Two expressions are said to be \textit{equivalent} if we can statically determine that both the expressions will have the same value during execution. 

The problem of Global Value Numbering(GVN) is how to extend the idea of local value numbering to detect redundant computations globally, within a program. An initial attempt on GVN can be found in Kildall \cite{kild}. In the \textit{structuring approach} of Kildall \cite{kild}, for every expression in the program, Kildall computes and maintains all its equivalent expressions, leading to exponential sized partitions. Most of the works that followed, tried to make the process more efficient  by means of special program representations and data structures \cite{awz, sumit, rosen, rks}. In general, we feel that there is a lack of clarity in the underlying concept of GVN and a lack of simplicity in the solutions.

In fact, in his implementation notes, Kildall suggested a \textit{value numbering approach} to ensure linear sized partitions \cite{kild}. An expression with value numbers as operands, called \textit{value expression}, is used to represent a set of expressions that are equivalent. At program points where multiple control flow paths merge, the equivalence information is computed by means of a \textit{confluence} operator. Confluence of equivalence classes involving \textit{value expression}s is a little bit tricky, and may be due to this reason the method did not draw much attention. Here, we make use of the concept of \textit{value expression} to devise a simple algorithm for GVN. 

We start with notations and definitions in Section 2. The algorithm for global value numbering is given in Section 3. This is followed by some comments on the algorithm in Section 4 and conclusions in Section 5. 

\section{Notations and Definitions}
\subsection{Program Representation}
The input to the algorithm is assumed to be a flow graph, with an empty entry node, denoted \textit{entry}, and an exit node, denoted \textit{exit}. Each node contains at most one assignment statement in three-address code\footnote {For simplicity, we do not consider other statements.}. Each assignment statement is of the form $x=e$, where $x$ is a variable and $e$ is an expression. An \textit{expression} is either a constant, a variable, or an expression of the form $y\ op\ z$ where $y$ and $z$ are variables or constants and $op$ is an operator. For a node $n$ in the flow graph, the input and output points of the node are denoted by $IN_n$ and $OUT_n$ respectively.   

\subsection{Expression-pool}
For a program point, the \textit{expression-pool} at that point denotes the set of expressions that are equivalent at that point. This is represented as a partition of expressions into equivalence classes. Each class in the pool has a \textit{value number}, denoted $v_i$, where $i$ is a positive integer. For convenience, we show the \textit{value number} of a class as the first element in it. As an example, $\{[v_1,\ a,\ x],\ [v_2,\ b,\ y]\}$ shows an \textit{expression-pool} with two equivalence classes. The value number  $v_1$ is assigned to $a$ and $x$ and value number $v_2$ is assigned to $b$ and $y$. For a node $n$ in the flow graph, we use $EIN_{n}$ and $EOUT_{n}$ to denote the \textit{expression-pool}s at $IN_n$ and $OUT_n$ respectively.

\subsection{Value Expression} For each expression of the form $x\ op\ y$, we can obtain a \textit{value expression}, by replacing the operands of the original expression by the corresponding \textit{value number}s. For example, suppose the expression-pool $\{[v_1,\ a,\ x],\ [v_2,\ b,\ y]\}$ reaches a node containing the statement $z=x+y$. Here, the \textit{value expression} of $x+y$ is $v_1+v_2$. Instead of the program expression $x+y$, we put its \textit{value expression} in the pool, with a new value number say $v_3$, to obtain the output pool, $\{[v_1,\ a,\ x],\ [v_2,\ b,\ y],\ [v_3,\ v_1+v_2,\ z]\}$.

We can see that the \textit{value expression} $v_1+v_2$, represents not just $x+y$, but the set of equivalent expressions $\{a+b,\ x+b,\ a+y,\ x+y\}$. The presence of $v_1+v_2$ in the pool indicates that an expression from this set is already computed, and this information is enough for detection of redundant computations. The interesting point to note is that a single binary \textit{value expression} can represent equivalence among any number of expressions of any length.  As an example, with respect to the above pool, the \textit{value expression} $v_1+v_3$ represents the expressions $a+z$, $x+z$, $a+(a+b)$, $a+(x+b)$, $x+(a+b)$, and so on. 

\section{Value Numbering: Algorithm}
The algorithm computes the expression-pool at every program point. For a node $n$ in the flow graph, if $EIN_{n}$ is given, then we can compute $EOUT_{n}$ by means of the \textit{transfer function} associated with node $n$. The points at which multiple control flow paths join are called \textit{confluence points}. For computing the \textit{expression-pool}s at such points, we define a \textit{confluence operator}.

\subsection{Transfer Function}
The \textit{transfer function} associated with a node $n$, denoted $f_n$, computes $EOUT_{n}$ using $EIN_{n}$, i.e. \textit {$EOUT_{n}=f_{n}(EIN_{n})$}. Algorithm 1 shows the transfer function for a node $n$ containing an assignment $x=e$, where \textit{x} is a program variable and \textit{e} is an expression\footnote{if $x$ occurs in $e$, we assume that the statement is split into $t=e$ followed by $x=t$ where $t$ is a new distinct temporary variable.}. The assignment can be considered as \textit{killing} all expressions involving the variable $x$ and \textit{generating} the new equivalence between $x$ and $e$. The effect of \textit{killing} expressions is achieved by removing $x$ from its class, say $C_i$. Now, if $C_i$ is a singleton with its value number, say $v_i$, as the only element in it, we delete the class $C_i$ and any \textit{value expression}s involving $v_i$ from the pool. The function $deleteSingletons(E)$ is assumed to do these steps repeatedly till no singleton classes remain in the pool $E$. 

The function $valueExp(e)$ returns the \textit{value expression} of $e$, if $e$ is of the form $x\ op\ y$, and returns $e$ itself otherwise. If $e$ is already assigned a value number in $EIN_{n}$, say $v_e$, then we put $x$ in the same class as that of $e$. Otherwise, a new class containing $x$ and $e$ is created together with a distinct \textit{value number} in it and this new class is  added to the output pool (if $e$ contains an operator, then instead of $e$, we add its value expression). 
\begin{algorithm}
\label{alg1}
 $E_{t}$ = $EIN_{n}$\;
{\bf if} ($x$ is in a class $C_x\in E_{t}$)\\
\hspace {0.2 in} {\bf then} remove $x$ from $C_x$\;
\hspace {0.55 in} $deleteSingletons(E_t)$\;
$e'=valueExp(e)$\;
 
{\bf if} ($e'$ is in a class $C_{e'}\in E_{t}$)\\
\hspace {0.2 in} {\bf then} add $x$ to $C_{e'}$\;
\hspace {0.2 in} {\bf else} create a new class $C_k$, with $x$ and $e'$ together with a new\\ \ \ \ \ \ \ \ \ \ \ \ \  value number $v_k$ in it, and add $C_k$ to $E_{t}$\;
$EOUT_{n} = E_{t}$\;
{return} $EOUT_{n}$\;
\caption{Computes $EOUT_{n} = f_{n}(EIN_{n})$, for a node $n$ containing the assignment $x=e$.}
\end{algorithm}

\subsection{Confluence Operator}
The \textit{expression-pool} at a confluence point should contain the sets of equivalent expressions common to all incoming pools. The common expressions that are explicitly present in the input pools can be obtained by a simple class-wise intersection of \textit{expression-pool}s. The hard part is obtaining the equivalence information based on the \textit{value expression}s in the incoming pools. 
\begin{figure}[ht]
\centering
\begin{tikzpicture}[scale=.75, transform shape, font=\sf\small,->,>=stealth',shorten >=1pt,auto,node distance=1.75cm,semithick]
  \tikzstyle{every state}=[fill=white,draw=white,text=black]

\begin{small}
  \node [circle]       (C) [distance=0.5cm, draw=black, fill=black] {};
 \node[rectangle, text width=6.5cm,minimum height=2mm] (A)[above left of=C, xshift=-2.25cm]   {\small {$E_1\colon \{[v_1,\ x,\ a],\ [v_2,\ y,\ b],\ [v_3,\ v_1+v_2,\ z]\}$}};
   \node[rectangle, text width=6.5cm,minimum height=2mm]         (B) [above right of=C, xshift=2.5cm] {\small{$E_2\colon \{[v_4,\ x,\ c],\ [v_5,\ y,\ d],\ [v_6,\ v_4+v_5,\ s]\}$}};
  \node[rectangle, text width=5.5cm,minimum height=2mm]         (D) [below of=C]       {\small{$E_3\colon \{[v_7,\ x\},\ [v_8,\ y],\  [v_9,\ v_7+v_8]\}$}};

 \path 
       (A) edge           (C)
	(B) edge           (C)
        (C)    edge           (D);
\end{small}
  \end{tikzpicture}
\caption{\small Computing confluence}
\end{figure}
In the example shown in Figure 1, we see two expression-pools, $E_1$ and $E_2$, reaching a confluence point. Let $E_3$ be the pool resulting after confluence. Since $x$ occurs in both the input pools $E_1$ and $E_2$, we put it in the output pool $E_3$. Since the value numbers of $x$ are different in the input pools, we assign a new distinct value number $v_7$ for the resulting class. Similarly, we put $y$ in  $E_3$ with new value number $v_8$. In $E_1$, the \textit{value expression} $v_1+v_2$ represents the set of equivalent expressions $x+y,\ x+b,\ a+y$, and $a+b$. In $E_2$, the \textit{value expression} $v_4+v_5$ represents the set of equivalent expressions $x+y,\ x+d,\ c+y$, and $c+d$. Here, we see a common expression $x+y$ represented by $v_1+v_2$ in $E_1$ and $v_4+v_5$ in $E_2$. Let us now devise a method to collect such common expressions by examining the \textit{value expression}s in the input pools. 

Consider the corresponding operands of the pair of \textit{value expression}s $v_1+v_2$ and $v_4+v_5$. We see a common element $x$ in the classes of $v_1$ and $v_4$ and a common element $y$ in the classes of $v_2$ and $v_5$. In other words, the intersection of the classes of $v_1$ and $v_4$ is non empty and also the intersection of the classes of $v_2$ and $v_5$ is non empty. This is enough to infer that there is a common expression represented by the two \textit{value expression}s. At confluence, the intersection of the classes of $v_1$ and $v_4$ results in a class with value number $v_7$ and the intersection of the classes of $v_2$ and $v_5$ results in a class with value number $v_8$. Hence the common expression $x+y$, after confluence, gets the \textit{value expression} $v_7+v_8$ and this can be added to $E_3$.

In general, let there be a class with value number $v_i$ and \textit{value expression} $v_{i1}+v_{i2}$ in $E_1$, and let there be a class with value number $v_j$ and \textit{value expression} $v_{j1}+v_{j2}$ in $E_2$. If the intersection of the classes of $v_{i1}$ and $v_{j1}$ results in a non empty class with \textit{value number} $v_{k1}$, and the intersection of the classes of $v_{i2}$ and $v_{j2}$ results in a non empty class with \textit{value number} $v_{k2}$, then  we can conclude that the pair of \textit{value expression}s, $v_{i1}+v_{i2}$ and $v_{j1}+v_{j2}$, represent a common expression $e$ in the two pools. The \textit{value expression} of $e$ after confluence is $v_{k1}+v_{k2}$ and this can be added to $E_3$.

An algorithm for computing the confluence of two expression-pools $E_i$ and $E_j$ is given in Algorithm 2. We use the symbol $\bigwedge$ to denote the confluence operation. The algorithm takes each pair of classes, $C_i \in E_i$ and $C_j \in E_j$, and finds the common expressions in $C_i$ and $C_j$ (either explicitly present or implicitly represented by \textit{value expression}s). The
\begin{algorithm}
\label{alg2}
$E_k=\Phi$ \;
{\bf foreach} pair of classes, $C_i \in E_i$ and $C_j \in E_j$\\
\hspace {0.2 in} $C_k=C_i \sqcap  C_j$\;
\hspace {0.2 in} {\bf if} ($C_k \ne \Phi$)\\
\hspace {0.4 in} {\bf then} add $C_k$ to $E_k$\;
$deleteSingletons(E_k)$\;
return $E_k$\;
\caption{Computing confluence of expression-pools, $E_i$ and $E_j$, i.e. $E_i\bigwedge E_j$.}
\end{algorithm}
operation of finding the common expressions in $C_i$ and $C_j$ can be considered as a special intersection and we denote it as $C_i \sqcap C_j$. Algorithm 3 shows the computation of $C_i \sqcap C_j$.
\begin{algorithm}[H]
\label{alg3}
$C_k=\Phi$\;
{\bf foreach} $e \in C_i \cap C_j$\\
\hspace {0.2 in}add $e$ to $C_k$\;
{\bf if} ($C_i$ and $C_j$ have different \textit{value expressions})\\
\hspace {0.2 in}{\bf then}\\
\hspace {0.4 in}\tcp {let $v_{i1}+v_{i2}$ and $v_{j1}+v_{j2}$ be the \textit{value expressions}}
\hspace {0.4 in}\tcp {in $C_i$ and $C_j$ respectively}
\hspace {0.4 in}$C_{k1}=C_{i1}\sqcap  C_{j1}$\;
\hspace {0.4 in}$C_{k2}=C_{i2}\sqcap  C_{j2}$\;
\hspace {0.4 in}{\bf if} ($C_{k1} \ne \Phi$ and $C_{k2} \ne \Phi$)\\
\hspace {0.6 in}{\bf then} add the \textit{value expression}  $v_{k1}+v_{k2}$ to $C_k$\;
{\bf if} ($C_k \ne \Phi$ and $C_k$ does not have a \textit{value number})\\
\hspace {0.2 in}{\bf then} add a new value number, say $v_k$, to $C_k$\;
return $C_k$\;
\caption{Computing $C_i \sqcap  C_j$. \newline Note: A class with value number $v_n$ is denoted by $C_n$ and vice-versa.}
 \end{algorithm}
\subsection{The Algorithm}
Algorithm 4 shows the main function for Global Value Numbering. $T$ is the $top$ element such that $E_i\bigwedge T=E_i$, for any expression-pool $E_i$. 
\begin{algorithm}
\label{alg4}
$EOUT_{entry}$ = $\Phi$\;
{\bf foreach} node $n \ne entry$ {\bf do} $EOUT_{n} = T$\; 
{\bf while} (changes to any $EOUT$ occur) {\bf do}\\
\hspace {0.2 in}\tcp {We mean changes in the equivalence information.}
\hspace {0.2 in}\tcp {The changes only in value numbers can be ignored.}
\hspace {0.2 in}{\bf foreach} node $n \ne entry$ {\bf do}\\
\hspace {0.4 in}$EIN_{n} = \bigwedge\limits_{p\in pred(n)}EOUT_{p}$\;
\hspace {0.4 in}$EOUT_{n}=f_{n}(EIN_{n})$\;
\caption{Computes $EIN_{n}$ and $EOUT_{n}$ for each node $n$.}
\end{algorithm}
For a node $n$, $pred(n)$ denotes the set of immediate predecessors of $n$, and $\bigwedge\limits_{p\in pred(n)}$ computes the confluence of expression-pools that reach the output of its predecessors. 
\section{Comments on the Algorithm}
\paragraph{Power of the Algorithm}
Figure 2 shows an example of redundancy detection which will demonstrate the power of the algorithm, especially that of the confluence operation.
\begin{figure}[ht]
\centering
\begin{tikzpicture}[scale=.75, transform shape, font=\sf\small,->,>=stealth',shorten >=1pt,auto,node distance=2.5cm,semithick]
 \node [circle]       (D) [distance=0.5cm, draw=black, fill=black] {};

\node[rectangle, text width=6.1cm,minimum height=2mm]   (A)[above left of=D,xshift=-1.7cm]     {\small {\hspace*{1 in}$c=a+b \newline \hspace*{1 in}e=c+z \newline \newline E_1\colon \{[v_1,\ x,\ a],\ [v_2,\ y,\ b],\ [v_3,\ v_1+v_2,\ c],$\\\hspace*{0.25 in}$\ [v_4,\ z],\ [v_5,\ v_3+v_4,\ e]\}$}};
  
  \node[rectangle, text width=6.1cm,minimum height=2mm]          (C) [above right of=D,xshift=2.2cm] {\small{\hspace*{1 in}$d=p+q \newline \hspace*{1 in}f=d+z\newline \newline E_2\colon \{[v_6,\ p,\ x],\ [v_7,\ q,\ y],\ [v_4,\ z],$\\\hspace*{0.3 in}$[v_8,\ v_6+v_7,\ d]$,$\ [v_9,\ v_8+v_4,\ f]\}$}};
  \node[rectangle, text width=9.1cm,minimum height=2mm]          (E) [below of=D,yshift=0.5cm]        {\small{$E_3\colon \{[v_{10},\ x\},\ [v_{11},\ y],\ [v_4,\ z], [v_{12},\ v_{10}+v_{11}],\ [v_{13},\ v_{12}+v_4] \}\newline \newline \hspace*{1.5 in} g=x+y$\newline \hspace*{1.5 in}$ h=g+z $}};
  \path 
        (C) edge              (D)
	(A) edge              (D)
(D) edge             (E);
  \end{tikzpicture}
\caption{\small Value Numbering to detect redundancy}
\end{figure}
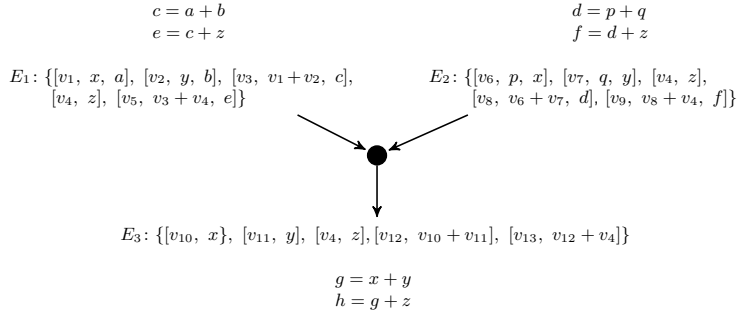
Let us use $C_i$ to denote a class with value number $v_i$. The \textit{value expression} $v_1+v_2$ in $C_3$ and $v_6+v_7$ in $C_8$ represent a common expression $x+y$. When we compute the confluence, $C_3 \sqcap C_8$ results in the class $C_{12}$ and the \textit{value expression} $v_{10}+v_{11}$ in it represents $x+y$. Another common expression is $(x+y)+z$, represented by the \textit{value expression}s in $C_5$ and $C_9$. The operation $C_5 \sqcap C_9$, results in $C_{13}$, whose \textit{value expression} $v_{12}+v_4$ represents $(x+y)+z$. After confluence, when we do value numbering of $g=x+y$, since $x+y$ maps to $v_{10}+v_{11}$, a \textit{value expression} in $E_3$, it is detected as redundant and $g$ gets \textit{value number} $v_{12}$. Similarly, the expression $g+z$ maps to $v_{12}+v_4$ and hence this is also detected as redundant.

\paragraph{A Comparison With Some of the GVN Algorithms}
In terms of power, our algorithm is as precise as Kildall's approach\cite{kild}. Alpern, Wegman, and Zadeck's (AWZ) algorithm \cite{awz} is an efficient algorithm for GVN, but is not as precise as Kildall's. The AWZ algorithm fails to detect the category of equivalences shown in Figure 2. The algorithm given by Gulwani and Necula \cite{sumit} does intersection of only those classes having at least one common variable. But as per our observation, intersection of all pairs of classes is required for detecting the kind of equivalences similar to that shown in Figure 2.  

\section{Conclusion}
An algorithm for Global Value Numbering is presented. The concept of \textit{value expression} enables a compact representation of equivalence and simplifies the computation of confluence. It may be noted that a single binary \textit{value expression} can represent equivalence among any number of expressions of any length. We feel the algorithm is simpler compared to that available in the literature. In terms of power, it is as precise as Kildall's approach.
\bibliography{mypapers}

\end{document}